\documentclass[runningheads]{llncs}
\usepackage{graphicx}
\usepackage{soul}
\usepackage{xcolor}
\usepackage{color,colortbl}
\definecolor{light}{rgb}{0.5, 0.5, 0.5}
\def\light#1{{\color{light}#1}} % light font

\usepackage{adjustbox}
\usepackage{booktabs} 
\usepackage{multirow}
\usepackage{float}
\usepackage{url}

\begin{document}
\title{Assessing the Effectiveness of LLMs in Android Application Vulnerability Analysis}
\author{Vasileios Kouliaridis \inst{1}\orcidID{} \and
Georgios Karopoulos \inst{1}\orcidID{} \and
Georgios Kambourakis \inst{2}\orcidID{}}

%\author{Blinded for review purposes}

%0000-0002-4233-5998 : V. Kouliaridis
%0000-0002-0142-7503 : G. Karopoulos
%0000-0001-6348-5031 : G. Kambourakis

%\authorrunning{V. Kouliaridis et al.}
\institute{European Commission, Joint Research Centre (JRC) 21027 Ispra, Italy \\ \and
Department of Information and Communication Systems Engineering, University of the Aegean, Karlovasi, 83200, Samos, Greece
}

\maketitle              

\begin{abstract}
%The increasing frequency of cyberattacks on Android applications necessitates a comprehensive understanding of the capabilities of large language models (LLMs) in identifying and mitigating security risks. The work at hand compares the ability of nine state-of-the-art LLMs to detect Android code vulnerabilities listed in the OWASP Mobile Top 10. To evaluate their performance, we created an open dataset of over 100 code samples, including obfuscated ones, that contain various key vulnerabilities and assessed each model's ability to identify them. Our analysis reveals the strengths and weaknesses of each model and identifies factors that contribute to their performance. Additionally, we provide valuable insights into context augmentation with Retrieval Augmented Generation for detecting Android code vulnerabilities, which in turn may propel the secure application development. Finally, while the reported findings regarding code vulnerability analysis show promise, they also reveal significant discrepancies among the different LLMs.

The increasing frequency of attacks on Android applications coupled with the recent popularity of large language models (LLMs) necessitates a comprehensive understanding of the capabilities of the latter in identifying potential vulnerabilities, which is key to mitigate the overall risk.
%Risk = Vulnerability x Threat
To this end, the work at hand compares the ability of nine state-of-the-art LLMs to detect Android code vulnerabilities listed in the latest Open Worldwide Application Security Project (OWASP) Mobile Top 10. Each LLM was evaluated against an open dataset of over 100 vulnerable code samples, including obfuscated ones, assessing each model's ability to identify key vulnerabilities. Our analysis reveals the strengths and weaknesses of each LLM, identifying important factors that contribute to their performance. Additionally, we offer insights into context augmentation with retrieval-augmented generation (RAG) for detecting Android code vulnerabilities, which in turn may propel secure application development. Finally, while the reported findings regarding code vulnerability analysis show promise, they also reveal significant discrepancies among the different LLMs.

\end{abstract}

\section{Introduction}

As mobile devices continue to proliferate, the need for secure software development practices remains still of high priority. The predominant Android platform has become a prime target for attackers and malware writers, seeking to exploit vulnerabilities in the vast cosmos of mobile applications~\cite{lookout2023}. The importance and volume of mobile vulnerabilities has led the Open Web Application Security Project (OWASP) to periodically publish a current, reputable list of the most prevalent vulnerabilities detected in mobile applications, namely OWASP Mobile Top 10.~\cite{owasp2024}. This list can serve as a key benchmark in assessing the performance of any tool in finding software vulnerabilities~\cite{kouliaridis2023}.

An emerging approach to detecting Android code vulnerabilities is the use of large language models (LLMs) for code analysis. Actually, the use of LLMs for code analysis is traced back to the early 2010s. That is, in 2013,
%Word2vec was created, patented, and published in 2013 by a team of researchers led by Mikolov at Google over two papers.
the introduction of Word2Vec~\cite{church2017word2vec}, a shallow neural network, marked the beginning of deep learning-based language models. That algorithm was capable of learning word embeddings (an encoding of the meaning of the word) from large datasets. In 2018,
%BERT was introduced in October 2018 by researchers at Google.
Google introduced Word2Vec's successor, a language model known as Bidirectional Encoder Representations from Transformers (BERT)~\cite{devlin2019bert}. BERT was designed to be bidirectionally trained, meaning it can learn information from both the left and right sides of a given text during training, therefore obtaining a better understanding of the context.

In the realm of code analysis, LLMs began to gain traction around 2017.
%Please recheck that the year provided (2017) is correct.
One of the early applications of LLMs in code analysis was code completion. Models like GPT-2~\cite{solaiman2019release}, fully released in Nov. 2019, were trained on a large corpus of source code data. By understanding the structure and context of the code, these models could predict the most likely code to follow a given input. In 2020, OpenAI~\cite{openai} introduced GPT-3~\cite{gpt3}, a significantly larger model with 175B parameters. This model showed improved capabilities in generating human-like text and was even able to generate code when given a task description. The ability of LLMs to analyze and understand code has also been demonstrated in recent studies~\cite{Wan2022,liu2023code}. Nevertheless, to the best of our knowledge, the literature lacks a comprehensive comparison of the ability of these models to detect Android code vulnerabilities so far.

The present work aims to fill this gap by comparing the ability of nine state-of-the-art LLMs to detect Android code vulnerabilities listed in the OWASP Mobile Top 10. Specifically, each model is evaluated regarding its performance in identifying key vulnerabilities against a dataset comprising snippets of vulnerable Android code. The assessment of each model is done through a combination of manual and automated evaluation methods. We additionally pinpoint the strengths and weaknesses of each LLM and provide insights into the factors that conduce to their performance. Overall, this study provides valuable insights into the use of LLMs for detecting mobile code vulnerabilities, thus contributing to the development of effective methods for secure mobile coding. The contributions of the paper are summarized as follows.

\begin{itemize}

\item We present a thorough comparative analysis on the capabilities and performance of nine leading LLMs, i.e., GPT 3.5, GPT 4, GPT 4 Turbo, Llama 2, Zephyr Alpha, Zephyr Beta, Nous Hermes Mixtral, MistralOrca, and Code Llama in identifying vulnerabilities residing in Android applications. The experiments conducted provide concrete evidence of the LLMs' capabilities for such tasks, also identifying the limitations per LLM. These insights are critical for anyone interested in understanding the trade-offs associated with each LLM.

\item We provide a comparison between the code analysis results as given by the nine LLMs against two well-known, publicly available static application security testing (SAST)  tools, namely, Bearer~\cite{bearer} and MobSFscan~\cite{mobsfscan}.

\item We examine the impact of context augmentation on LLMs and contribute a set of guidelines regarding the selection and fine-tuning of LLMs towards enhancing the security posture of Android code.

\item We offer an open dataset to the community for driving research in this field forward. 

\end{itemize}

The remainder of this paper is structured as follows. The next section presents previous work on the use of LLM for code vulnerability analysis. Section~\ref{S:methodology} details our methodology, while the results per LLM are given in section~\ref{S:results}. The last section concludes and proposes some lines for future research.

\section{Previous work}
\label{S:previous_work}

In recent years, LLMs have gained significant attention in the field of cybersecurity for their potential to provide assistance in various domains, including vulnerability detection, penetration testing, and security analysis. 
State-of-the-art surveys such as~\cite{AlHawawreh2023ChatGPT,Yao24} and~\cite{Gupta23}, as well as a more recent but not yet peer-reviewed study~\cite{motlagh2024large}, provide comprehensive overviews of the current state and potential future applications of LLMs in cybersecurity. These works analyze the challenges, practical implications, and future research directions to exploit the full potential of these models in ensuring cyber resilience.
The rest of this section will focus on literature dealing with software vulnerability analysis using LLMs. This includes works that have already been peer-reviewed, as well as more recent research that has been self-archived for the sake of completeness.

In~\cite{thapa22}, transformer-based LLMs are evaluated in the task of code vulnerability detection. The authors evaluate such LLMs, including BERT, DistilBERT, CodeBERT, GPT-2 and Megatron, against C/C++ source code snippets from two publicly available datasets. The results showed that LLMs perform well in software vulnerability tasks; indicatively, the best scoring model, GPT-2, had an F1-score above 95\% in all tests. 
In the context of software engineering,~\cite{Liu2023} investigates the use of in-context learning to improve the ability of LLMs to detect software vulnerabilities, showcasing the adaptability of LLMs to learn from context-specific examples. The authors use code retrieval to search for code snippets that are similar to the examined code and feed them to the LLM together with the examined code and its analysis. Their experimental results show that this approach has better performance than the original GPT model.

Another set of works, adds verification in the vulnerability detection process.
An empirical study of using LLMs for vulnerability assessment in software was conducted in~\cite{Purba2023}. The authors used four well-known pre-trained LLMs to identify vulnerabilities in two labeled datasets, namely code gadgets and CVEfixes, and static analysis as a reference point. The used LLMs include GPT-3.5, Davinci and CodeGen, and the analysis was limited to two kinds of vulnerabilities: SQL injections and buffer overflows. The study concluded that LLMs do not perform well at detecting vulnerabilities, presenting high false-positive rates, but could complement and improve the traditional static analysis process.
Concerns about the safe use of code assistants are addressed in~\cite{sandoval2023lost}. In this case, LLMs are used to produce code which is then assessed manually and using static analysis. This study provides empirical insights into how developers interact with LLMs, underscoring the importance of user awareness to mitigate security risks associated with assisted code generation.

%\kar{assesses Python code, LLM used for code generation, no LLM use in code assessment, no peer-review, 2023}
%The work by \cite{siddiq2023generate} represents a crucial step in evaluating the security robustness of code generated by LLMs. This study introduced SALLMS, a systematic approach to assess the security implications of LLM-generated code, highlighting the potential risks and the need for rigorous validation mechanisms.

Moving to non peer-reviewed works, the work of~\cite{liu2023harnessing} delves into the application of LLMs in static binary taint analysis, demonstrating how these models can assist in vulnerability inspection of binaries. A binary is first disassembled and decompiled, and an LLM is used to identify security sensitive functions that may contain vulnerabilities, as well as candidate dangerous flows. In the last phase, the LLM combines the previous results to produce a vulnerability report for the examined binary.
%\kar{part of it uses LLM, LLM is pre-trained, no peer-review, 2023}
The authors of~\cite{wang2023defecthunter}  propose DefectHunter, a vulnerability detection mechanism that combines various technologies, including LLMs. Its architecture has three main building blocks: a tool for extracting structural information from code snippets, a pre-trained LLM for generating semantic information, and a Conformer mechanism to identify vulnerabilities from the previously extracted structural and semantic data.

%\kar{code vulnerability using LLMs, technical report, no peer-review, 2023}
The authors of~\cite{cheshkov2023evaluation} evaluated ChatGPT and GPT-3 in detection of Common Weakness Enumeration (CWE) vulnerabilities contained in code. Using a custom real-world dataset with Java files from open GitHub repositories, they concluded that the detection capabilities of the aforementioned models are limited.
%\kar{vulnerability assessment using LLMs, comparison of results with static analysis, no peer-reviewed, 2023}
In~\cite{noever2023large}, an empirical study of the potential of LLMs for detecting software vulnerabilities is presented. The authors tested 129 code samples from various GitHub repositories, written in eight different languages, and their results showed that GPT-4 identified around four times more vulnerabilities than traditional, rule-based, static code analysis tools. In addition, the LLMs were asked to provide fixes for the identified vulnerabilities. The models used include GPT-3 and GPT-4.

Apart from generic code, LLMs have been used for detecting vulnerabilities in smart contracts.
%\kar{evaluation framework for vulnerability detection systems using LLMs, focus is on smart contracts , no peer reviewed, 2024}
LLM4Vuln~\cite{sun2024llm4vuln} is an evaluation framework for vulnerability detection systems based on LLMs, focusing on smart contract vulnerabilities. The difference from other similar works is that, instead of benchmarking the performance of LLMs in vulnerability detection, the authors evaluate the vulnerability reasoning capabilities of each model.  
%\kar{vulnerability analysis of smart contracts, no peer-review, 2023}
Similarly, the authors of~\cite{hu2023large} proposed GPTLens, a framework for detecting vulnerabilities in smart contracts using LLMs. GPTLens takes a different approach from the traditional one-stage detection in order to decrease false positives. The detection process is broken down in two steps, where the LLM takes two different roles: auditor and critic. As an auditor, the LLM provides a large range of vulnerabilities for the examined contract, whereas as a critic it verifies the claims produced in the first step. The performed experiments show that GPTLens presents improved results over the single-stage vulnerability detection. 

% \kar{not related to vulnerability detection} Additionally, the authors of~\cite{NEURIPS2020} showed that RAG models obtain state-of-the-art results without retraining the models, validating its effectiveness.

%\kar{using LLMs to repair already identified security bugs in source code}
%The study of \cite{Pearce2023} explores the capacity of LLMs to not only detect but also repair software vulnerabilities in a zero-shot setting. This research is pivotal in understanding the potential for LLMs to contribute to the automatic remediation of security flaws.

%\kar{not software vulnerability detection but penetration testing}
%Several empirical studies and tool development efforts, such as , ,  \cite{deng2023pentestgpt}, contribute to the growing evidence that LLMs can be effective in automating penetration testing tasks.

%\kar{application of LLMs to system-on-chip (SoC) security}
%\cite{saha2023llm}

\section{Methodology}
\label{S:methodology}

%This section describes the methodology of our experiments, namely the new dataset, the selection of the state-of-the-art LLMs, and the evaluation process.

This section details our methodology, including the creation of the benchmark dataset, the selection of LLMs, and the evaluation process.

\subsection{Dataset}
\label{SS:Dataset}

Also with reference to Section~\ref{S:previous_work}, to our knowledge, there is no publicly available dataset containing vulnerable Android code covering each one of the OWASP Mobile Top 10 vulnerabilities. The most relevant dataset to our study is LVDAndro~\cite{LVDAndro}, which however is labelled based on CWE. Additionally, since LVDAndro was created using actual Android applications, it contains a significant proportion of non-vulnerable code. In view of this shortage, for the needs of our experiments, we created a new dataset coined \textit{Vulcorpus}~\cite{vulcorpus2024} containing 100 pieces of vulnerable code. It is important to note that the term ``piece of code'', hereafter called \textit{sample}, refers to a part of an application, not its full codebase. All the samples were written in Java by exploiting common insecure coding practices, e.g., logging private information, not filtering input/objects, etc., and target the Android OS. However, obviously, the same vulnerabilities apply to other mobile platforms, say, iOS.

More specifically, Vulcorpus contains 10 samples for each of the OWASP Mobile Top-10 vulnerabilities of 2024, which are briefly explained in subsection~\ref{SS:List:of:Vulnerabilities}. Every sample exhibits one or maximum two interrelated vulnerabilities, while one or two of these samples per vulnerability category are obfuscated using the well-known renaming technique. Half of the samples per vulnerability contain code comments regarding the specific vulnerability. 
%Vulcorpus is publicly available at Github~\cite{vulcorpus2024}. 
Moreover, to assess each LLM in detecting privacy-invasive code, we created three more samples which perform risky actions without asking the user for confirmation. These actions are:

\begin{itemize}
    \item Get the precise location of the device through the ``android.permission.ACCESS\_FINE\_LOCATION'' permission, and directly share the latitude and longitude over the Internet via API. According to the Android API~\cite{fine_loc}, this permission has a ``dangerous'' protection level, namely it may give the requesting application access to user's private data, among others.
    \item Capture an image via the ``ACTION\_IMAGE\_CAPTURE'' intent~\cite{image_intent}%, controlling the storing location with an extra (Bundle) ``EXTRA\_OUTPUT''
    , and subsequently attempt to share the captured image file via API.
    \item Open local documents through the ``ACTION\_OPEN\_DOCUMENT'' intent~\cite{open_docs}, and attempt to send them to a remote host via API. 

\end{itemize}

The latter three samples are also available at~\cite{vulcorpus2024} along with Vulcorpus.
%\gk{Did you write all the samples or you collected them from one or more repos?} \vk{I wrote all of them.} \gk{Are these samples current or pertain to obsolete anymore coding practices that have been proved to lead to vulnerabilities?} \vk{I do not use any oudated libraries, i mostly use vulnerable practices, such as logging private information, not filtering input/objects, etc. these practices exist in the wild.} \gk{Were the code samples obfucated?} \vk{Only 1-2 in each category.} \gk{Why not including safe code samples (the normal class) in the dataset for possible false positives?} \vk{I did not think it was necessary as this experiment doesn't include a training process}.

\subsection{List of vulnerabilities}
\label{SS:List:of:Vulnerabilities}

This subsection briefly delineates each vulnerability contained in the current OWASP Mobile Top 10 list. For more details regarding each vulnerability, the reader is referred to~\cite{owasp2024}. It is important to note that the list differs from its 2016 version, given that four vulnerabilities contained in the 2016 list have been replaced with new ones in the current list. The reader should also keep in mind that while some categories of vulnerabilities, say, M5 are straightforward, others might be more complicated for LLMs to understand, such as the M7.

\noindent \textbf{Improper credential usage (M1)}: Poor credential management can lead to severe security issues, namely, unauthorized users may be able to gain access to sensitive information or administrative functionalities within the mobile app or its backend systems. This in turn leads to data breaches and fraudulent activities.

\noindent \textbf{Inadequate supply chain security (M2)}: By exploiting vulnerabilities in the mobile supply chain, attackers may be able to manipulate application functionality. For example, they can insert malicious code into the mobile application's codebase or libraries~\cite{xzutils}, as well as modify the code during the application's build process to introduce backdoors, spyware, or other type of malware. The attacker can also exploit vulnerabilities in third-party software libraries, software development kits (SDKs), or hard-coded credentials to gain access to the mobile app or the backend servers. Overall, this type of vulnerabilities can lead to unauthorized data access or manipulation, denial of service, or complete takeover of the mobile application or device.

\noindent \textbf{Insecure authentication/authorization (M3)}: Poor authorization could lead to the destruction of systems or unauthorized access to sensitive information, while poor authentication results in the inability to identify the user making an action request, leading to the inability to log or audit user activity. This situation makes it difficult to detect the source of an attack, understand any underlying exploits, or develop strategies to prevent future attacks. Obviously, authentication failures are tightly coupled to authorization failures; when authentication controls fail, authorization cannot be performed. That is, if an attacker can anonymously execute sensitive functionality, it indicates that the underlying code is not verifying the user's permissions, highlighting failures in both authentication and authorization controls.

\noindent \textbf{Insufficient input/output validation (M4)}: A mobile application that does not adequately validate and sanitize data from external sources, like user inputs or network data, is susceptible to a range of attacks, including SQL injection, command injection, and cross-site scripting. Insufficient output validation can also lead to data corruption or presentation vulnerabilities, possibly allowing the malicious actor to inject harmful code or manipulate sensitive information shown to the users.

\noindent \textbf{Insecure communication (M5)}: Modern mobile applications typically communicate with one or more remote servers. This renders user data susceptible to interception and modification, if they are transmitted in plaintext or using an outdated encryption protocol.

\noindent \textbf{Inadequate privacy controls (M6)}: Privacy controls aim to safeguard Personally Identifiable Information (PII), including names and addresses, credit card details, emails, and information related to health, religion, sexuality, and political opinions. This sensitive information can be used to impersonate the victim for fraudulent activities, misuse their payment data, blackmail them with sensitive information, or harm them by destroying or manipulating sensitive data.

\noindent \textbf{Insufficient binary protections (M7)}: The application's binary may hold valuable information, such as commercial API keys or hard-coded cryptographic secrets. Furthermore, the code within the binary itself could be valuable, for instance, containing critical business logic or pre-trained AI models. In addition to gathering information, attackers may also manipulate app binaries to gain access to paid features for free or to bypass other security controls. In the worst-case scenario, popular apps could be altered to include malicious code and then distributed through third-party app stores or under a new name to deceive unsuspecting users.

\noindent \textbf{Security misconfiguration (M8)}: These occur when security settings, permissions, or controls are improperly configured, leading to vulnerabilities and unauthorized access.

\noindent \textbf{Insecure data storage (M9)}: Such vulnerabilities may stem from weak encryption, insufficient data protection, insecure data storage mechanisms, and improper handling of user credentials.

\noindent \textbf{Insufficient cryptography (M10)}: The use of obsolete cryptographic suites, primitives, or cryptographic practices may lead to loss of data confidentiality, integrity, and inability to impose source authentication among others. Typical repercussions include data decryption, manipulation of cryptographic processes, leak of encryption keys, etc. 

%\subsection{Selection of state-of-the-art LLMs}
\subsection{Selection of LLM}
\label{SS:LLMs_collection}

%[TODO: extend subsection and add citations]
%We selected nine well-known LLMs namely GPT 3.5, GPT 4, GPT 4 Turbo, Llama 2, and Zephyr Alpha, Zephyr Beta, Nous Hermes Mixtral, MistralOrca, and Code Llama to evaluate their effectiveness on detecting Android vulnerabilities. These models have been pre-trained on large amounts of text data, including code, and have demonstrated performance in various software engineering tasks, including code analysis and have shown promising results. More specifically:

For the purposes of our experiments, nine contemporary, well-known LLMs were chosen: three commercial models, i.e., GPT 3.5, GPT 4, and GPT 4 Turbo, and six open source models, i.e., Llama 2, Zephyr Alpha, Zephyr Beta, Nous Hermes Mixtral, MistralOrca, and Code Llama. According to their documentation, these models have been pre-trained on large amounts of text data, including code, having demonstrated performance in various software engineering tasks, including code analysis. That is, their ability to understand code syntax and semantics makes them well-suited for identifying vulnerabilities residing in code. Additionally, their large size and diverse training data make them less likely to overfit to a specific codebase. A succinct description of each LLM is given below.

\begin{itemize}

    \item GPT 3.5 (gpt-35-turbo version Nov. 2023)~\cite{brown2020language}: It is a powerful language model that has been pre-trained on a large corpus of text data, including code. It has demonstrated performance in various natural language processing (NLP) tasks and has been used for code analysis tasks such as code completion, code search, and code summarization.
    
    \item GPT 4~\cite{gpt4},~\cite{openai2024gpt4}: It is the newest version of GPT being pre-trained on an even larger corpus of text data, including code. It has demonstrated improved performance over GPT 3.5 in various NLP tasks and has been used for code analysis, including code review and repair.
    
    \item GPT 4 Turbo (gpt-4-1106): It is a variant of GPT 4, been specifically designed for tasks that require faster inference times, such as code analysis. It has been pre-trained on the same large corpus of text data as GPT 4, optimized for faster performance.
    
    \item Llama 2 (Llama-2-70b-chat)~\cite{touvron2023Llama}: This LLM has been pre-trained on a diverse set of text data, including code. It has demonstrated performance in various NLP tasks, also been exploited for code analysis, including code summarization and code search. 
    
    \item Zephyr Alpha (zephyr-7b-alpha)~\cite{tunstall2023zephyr}: It is  pre-trained on a huge corpus of text data from diverse sources, including books, articles, and websites. This model has been fine-tuned with a mix of publicly available and synthetic datasets on top of Mistral LLM. Despite its small size (7B parameters), it potentially shows a performance comparable to several models with a number of parameters in the range of 20-30B.
    
    \item Zephyr Beta (zephyr-7b-beta)~\cite{tunstall2023zephyr}: This model has been fine-tuned with a mix of publicly available and synthetic datasets on top of Mistral LLM. It is the successor of Zephyr Alpha, therefore considered significantly more powerful than its predecessor. Based on its documentation, it is fast and competent, showing a performance comparable to the best open-source models, having around 70B parameters.
    
    \item Nous Hermes Mixtral (nous-hermes-2-mixtral-8x7b-dpo)~\cite{NousHermesMixtral}: It is one of the most powerful open-source models available, comprising a fine-tuned version of Mixtral base model.
    
    \item MistralOrca (mistral-7b-openorca~\cite{lian2023mistralorca1},~\cite{mukherjee2023orca},~\cite{longpre2023flan}: It has been fine-tuned with Open-Orca datasets on top of Mistral LLM. Despite its small size, it outperforms Llama 2 13B, showing a performance comparable to several models with a number of parameters in the range of 20-30B.
    
    \item Code Llama~\cite{codeLlama}: It is a special version of Llama 2, tailored specifically for coding applications. This specialized version has been refined through extensive additional training on code-focused data, with prolonged exposure to relevant datasets. The result is a tool with alleged superior coding proficiency that builds upon the foundation of Llama 2. More specifically, Code Llama can generate code and create explanations about code in response to prompts in both programming and natural language. Its capabilities extend to assisting with code completion and troubleshooting code errors. Furthermore, Code Llama is versatile, supporting a broad array of widely-used programming languages, including Python, C++, Java, PHP, JavaScript, C Sharp, and Bash. In this work, we examine the smallest pre-trained model, namely, the 7B version.
    In addition, for this LLM, in a separate run, we employed LlamaIndex~\cite{LlamaIndex} to improve the detection capabilities of Code Llama. LlamaIndex is a data framework for LLM-based applications, enhancing them with additional contextual data. This context augmentation technique is called Retrieval-Augmented Generation (RAG) and can be used to address the restrictions of LLMs by giving them access to contextual, current data. For the RAG process, we used the 50\% of Vulcorpus, i.e., only the samples that contain code comments regarding the specific vulnerability. Android's application quality and security guidelines and code examples~\cite{AndroidSecGuide} were also added as input to the RAG, along with information on each vulnerability from the OWASP website~\cite{owasp2024}.
    
\end{itemize}

\subsection{Evaluation process}
\label{SS:Evaluation:Process}

All nine pre-trained LLMs listed in subsection~\ref{SS:LLMs_collection}, except Code Llama, run on the \textit{GPT@JRC} platform, a system developed by the European Commission's Joint Research Centre (JRC). Code Llama was run on a local computer with an M2 processor and 16GB unified memory. Each LLM was fed with Vulcorpus for comparing its performance on identifying potential vulnerabilities and proposing code improvements. To this end, as detailed in Section~\ref{S:results}, we use a simple scoring system to present (a) the number of vulnerabilities each LLM was able to detect, and (b) if the LLM proposed valid suggestions for possibly fixing the vulnerability. Both these partial scores have a maximum value of 10/10 per vulnerability category, i.e., one point for each piece of vulnerable code the LLM was able to detect and annotate. It is important to note that the input or question given to each LLM has a major effect on its output. For our study, each LLM was queried as follows: ``Check if there are any security issues in the following code; if there are, explain the issue''.

As previously mentioned, the LLMs used in this work are pre-trained. This means that the associated libraries, possibly needed by each code sample but not included in the input, cannot be analyzed. This mostly affects the analysis regarding the M2 vulnerability. Therefore, to evaluate LLMs against M2, instead of Java code, we used 10 libraries with known vulnerabilities as input. These libraries, also included in Vulcorpus for reasons of reproducibility, were published before the training date of each LLM.

At a final stage, as detailed in Section~\ref{S:results}, the results of each LLM were compared and crosschecked against those produced by two well-known SAST tools, namely Bearer~\cite{bearer} and MobSFscan~\cite{mobsfscan}. Bearer is a static application security testing tool, which uses built-in rules covering the OWASP Top 10 and Common Weakness Enumeration (CWE) Top 25. MobSFscan is a static analysis tool that uses MobSF's~\cite{mobsf} security rules and can find insecure code patterns in Android or iOS source code. Finally, we also assessed the performance of each LLM in detecting privacy-invasive behaviors, using the three samples detailed in subsection~\ref{SS:Dataset}. The output was rated using three categories: not privacy-invasive, (b) potentially privacy-invasive, and (c) privacy-invasive.

\section{Results}
\label{S:results}

Tables~\ref{T:va_results} and~\ref{T:va_vuln} recapitulate the results for each LLM. Particularly, each line of Table~\ref{T:va_results} indicates if the specific model \textit{Detected} the vulnerability (denoted with the letter ``D''), and if it explained the situation and provided a valid solution for \textit{Improving} the code (denoted with the letter ``I''). Actually, the ``I'' aspect is a key factor in evaluating each LLM (also against each other), as this is the sole indicator of whether the LLM actually ``perceives'' the security issue.

\begin{table}
\centering
\caption{Vulnerability analysis results. The letters ``D'' and ``I'' stand for the number of vulnerable samples detected and the number of vulnerable samples for which the LLM suggested improvements, respectively. Top scores per vulnerability are in boldface. The asterisk exhibitor stands for Code Llama without RAG. }
\label{T:va_results}
%\begin{adjustbox}{max width=\textwidth}
\begin{tabular}{p{2cm}rrrrrrrrrrrrrrrrrrrrrrr}
\toprule
 
\multicolumn{1}{c|}{\multirow{2}{*}{\textbf{LLM}}} & \multicolumn{2}{|c|}{\textbf{M1}} & \multicolumn{2}{|c|}{\textbf{M2}} & \multicolumn{2}{|c|}{\textbf{M3}} & \multicolumn{2}{|c|}{\textbf{M4}} & \multicolumn{2}{|c|}{\textbf{M5}} & \multicolumn{2}{|c|}{\textbf{M6}} & \multicolumn{2}{|c|}{\textbf{M7}} & \multicolumn{2}{|c|}{\textbf{M8}} & \multicolumn{2}{|c|}{\textbf{M9}} & \multicolumn{2}{|c|}{\textbf{M10}} & \multicolumn{2}{|c|}{\textbf{Mean}}\\ \cline{2-23}

\multicolumn{1}{c|}{} & \multicolumn{1}{|c|}{\textbf{D}} & \multicolumn{1}{|c|}{\textbf{I}} & \multicolumn{1}{|c|}{\textbf{D}} & \multicolumn{1}{|c|}{\textbf{I}} & \multicolumn{1}{|c|}{\textbf{D}} & \multicolumn{1}{|c|}{\textbf{I}} & \multicolumn{1}{|c|}{\textbf{D}} & \multicolumn{1}{|c|}{\textbf{I}} & \multicolumn{1}{|c|}{\textbf{D}} & \multicolumn{1}{|c|}{\textbf{I}} & \multicolumn{1}{|c|}{\textbf{D}} & \multicolumn{1}{|c|}{\textbf{I}} & \multicolumn{1}{|c|}{\textbf{D}} & \multicolumn{1}{|c|}{\textbf{I}} & \multicolumn{1}{|c|}{\textbf{D}} & \multicolumn{1}{|c|}{\textbf{I}} & \multicolumn{1}{|c|}{\textbf{D}} & \multicolumn{1}{|c|}{\textbf{I}} & \multicolumn{1}{|c|}{\textbf{D}} & \multicolumn{1}{|c|}{\textbf{I}} & \multicolumn{1}{|c|}{\textbf{D}} & \multicolumn{1}{|c|}{\textbf{I}}\\
\hline

GPT-3.5 & 3 & \cellcolor[gray]{0.8}3 & \textbf{7} & \cellcolor[gray]{0.8}N/A & 2 & \cellcolor[gray]{0.8}3 & 2 & \cellcolor[gray]{0.8}3 & 8 & \cellcolor[gray]{0.8}6 & 3 & \cellcolor[gray]{0.8}5 & 5 & \cellcolor[gray]{0.8}4 & 3 & \cellcolor[gray]{0.8}5 & 4 & \cellcolor[gray]{0.8}6 & 5 & \cellcolor[gray]{0.8}2 & 4.2 & \cellcolor[gray]{0.8}4.1 \\ \hline
GPT-4 & \textbf{10} & \cellcolor[gray]{0.8}\textbf{10} & 0 & \cellcolor[gray]{0.8}N/A & 6 & \cellcolor[gray]{0.8}7 & 6 & \cellcolor[gray]{0.8}\textbf{8} & 5 & \cellcolor[gray]{0.8}\textbf{10} & \textbf{10} & \cellcolor[gray]{0.8}\textbf{10} & 6 & \cellcolor[gray]{0.8}\textbf{9} & 7 & \cellcolor[gray]{0.8}\textbf{10} & 8 & \cellcolor[gray]{0.8}\textbf{10} & 9 & \cellcolor[gray]{0.8}9 & 6.7 & \cellcolor[gray]{0.8}\textbf{9.2} \\ \hline
GPT-4 TURBO & 4 & \cellcolor[gray]{0.8}5 & 0 & \cellcolor[gray]{0.8}N/A & 3 & \cellcolor[gray]{0.8}5 & 5 & \cellcolor[gray]{0.8}\textbf{8} & 8 & \cellcolor[gray]{0.8}9 & 6 & \cellcolor[gray]{0.8}8 & 4 & \cellcolor[gray]{0.8}4 & 7 & \cellcolor[gray]{0.8}\textbf{10} & 7 & \cellcolor[gray]{0.8}9 & 6 & \cellcolor[gray]{0.8}8 & 5 & \cellcolor[gray]{0.8}7.3 \\ \hline
Nous Hermes Mixtral & 1 & \cellcolor[gray]{0.8}3 & 6 & \cellcolor[gray]{0.8}N/A & 1 & \cellcolor[gray]{0.8}3 & \textbf{9} & \cellcolor[gray]{0.8}5 & 6 & \cellcolor[gray]{0.8}8 & 8 & \cellcolor[gray]{0.8}8 & 7 & \cellcolor[gray]{0.8}3 & \textbf{9} & \cellcolor[gray]{0.8}9 & 8 & \cellcolor[gray]{0.8}\textbf{10} & 7 & \cellcolor[gray]{0.8}7 & 6.2 & \cellcolor[gray]{0.8}6.2 \\ \hline
Mistral Orca & 9 & \cellcolor[gray]{0.8}9 & 0 & \cellcolor[gray]{0.8}N/A & 2 & \cellcolor[gray]{0.8}2 & 4 & \cellcolor[gray]{0.8}4 & 5 & \cellcolor[gray]{0.8}5 & 0 & \cellcolor[gray]{0.8}0 & 3 & \cellcolor[gray]{0.8}4 & 0 & \cellcolor[gray]{0.8}1 & \textbf{10} & \cellcolor[gray]{0.8}\textbf{10} & 4 & \cellcolor[gray]{0.8}3 & 3.7 & \cellcolor[gray]{0.8}4.2 \\ \hline
Zephyr Alpha & 0 & \cellcolor[gray]{0.8}0 & 3 & \cellcolor[gray]{0.8}N/A & 6 & \cellcolor[gray]{0.8}6 & 5 & \cellcolor[gray]{0.8}5 & \textbf{10} & \cellcolor[gray]{0.8}\textbf{10} & 7 & \cellcolor[gray]{0.8}8 & 2 & \cellcolor[gray]{0.8}2 & 7 & \cellcolor[gray]{0.8}7 & 3 & \cellcolor[gray]{0.8}\textbf{10} & \textbf{10} & \cellcolor[gray]{0.8}\textbf{10} & 5.3 & \cellcolor[gray]{0.8}6.4 \\ \hline
Zephyr Beta & 0 & \cellcolor[gray]{0.8}8 & 0 & \cellcolor[gray]{0.8}N/A & \textbf{9} & \cellcolor[gray]{0.8}9 & 8 & \cellcolor[gray]{0.8}\textbf{8} & \textbf{10} & \cellcolor[gray]{0.8}\textbf{10} & 0 & \cellcolor[gray]{0.8}8 & 3 & \cellcolor[gray]{0.8}0 & 5 & \cellcolor[gray]{0.8}4 & \textbf{10} & \cellcolor[gray]{0.8}0 & 9 & \cellcolor[gray]{0.8}9 & 5.4 & \cellcolor[gray]{0.8}6.2 \\ \hline
Llama 2 & 0 & \cellcolor[gray]{0.8}0 & 0 & \cellcolor[gray]{0.8}N/A & 0 & \cellcolor[gray]{0.8}\textbf{10} & 4 & \cellcolor[gray]{0.8}5 & \textbf{10} & \cellcolor[gray]{0.8}0 & 6 & \cellcolor[gray]{0.8}6 & 0 & \cellcolor[gray]{0.8}0 & 0 & \cellcolor[gray]{0.8}0 & 4 & \cellcolor[gray]{0.8}4 & 6 & \cellcolor[gray]{0.8}6 & 3 & \cellcolor[gray]{0.8}3.4 \\ \hline
Code Llama* & 9 & \cellcolor[gray]{0.8}5&3&\cellcolor[gray]{0.8}N/A&\textbf{9}&\cellcolor[gray]{0.8}4&8&\cellcolor[gray]{0.8}4&\textbf{10}&\cellcolor[gray]{0.8}5&8&\cellcolor[gray]{0.8}5&\textbf{9}&\cellcolor[gray]{0.8}4&\textbf{9}&\cellcolor[gray]{0.8}4&7&\cellcolor[gray]{0.8}7&9&\cellcolor[gray]{0.8}6 & \textbf{8.1} & \cellcolor[gray]{0.8}4.9 \\
\bottomrule

\end{tabular}
%\end{adjustbox}
\end{table}

\begin{table}
\centering
\caption{Results per LLM regarding privacy-invasive actions. N: not privacy-invasive, P: potentially privacy-invasive, Y: privacy-invasive.}
\label{T:va_vuln}
%\begin{adjustbox}{max width=\textwidth}
\begin{tabular}{lccc}
\toprule

\textbf{LLM} & \textbf{Location} & \textbf{Camera} & \textbf{Files} \\ 
\hline

GPT 3.5 & \textbf{Y} & P & P \\ \hline
GPT 4  & \light{N} & P &\light{N} \\ \hline
GPT 4 Turbo  & P & \textbf{Y} & P \\ \hline
Nous Hermes Mixtral  & \textbf{Y} & P & P \\ \hline
MistralOrca  & \light{N} & \light{N} & \light{N} \\ \hline
Zephyr Alpha  & \textbf{Y} & P & \textbf{Y} \\ \hline
Zephyr Beta  & \textbf{Y} & P &\light{N} \\ \hline
Llama 2  & P & \textbf{Y} & \light{N} \\ \hline
Code Llama  &\light{N} & \light{N} & \textbf{Y} \\ \hline
Code Llama + RAG  & \light{N} & \textbf{Y} & P \\
\bottomrule

\end{tabular}
%\end{adjustbox}
\end{table}

Overall, with reference to Table~\ref{T:va_results}, the best performers in terms of total vulnerabilities detected, are Code Llama (81/100), GPT 4 (67/100), Nous Hermes Mixtral (62/100), Zephyr Beta (54/100), and Zephyr Alpha (53/100), followed by GPT 4 TURBO (50/100), GPT 3.5 (42/100), MistralOrca (37/100), and Llama 2 (30/100). On the other hand, the best performers, in terms of total code improvement suggestions, are GPT 4 (83/90), GPT 4 Turbo (66/90), Zephyr Alpha (58/90), Zephyr Beta (56/90), and Nous Hermes Mixtral (56/90), followed by Code Llama (44/90), MistralOrca (38/90), GPT 3.5 (37/90), and Llama 2 (31/90). Overall, GPT 4 poses as the top performer, considering a composite score of high ``D'' and high ``I''. On the other hand, LLMs like Code Llama, which do identify the correct vulnerability, but fail to provide corrections or suggestions regarding the problematic lines of code may indicate an insufficiently trained model for this type of analysis.

When looking at each vulnerability individually, GPT 4 achieved a perfect score for M1 and M6, MistralOrca for M9, Zephyr alpha for M5 and M10, Zephyr beta for M5 and M9, and Llama 2 and Code Llama for M5. Regarding the rest of the vulnerabilities, namely, M2, M3, M4, M7, and M8, the best performers were GPT 3.5 (7/10), Zephyr Beta and Code Llama (9/10), Nous Hermes Mixtral (9/10), Code Llama (9/10), and Nous Hermes Mixtral and Code Llama (9/10), respectively.
Concerning M2, recall from subsection~\ref{SS:Evaluation:Process} that it was tested using 10 vulnerable libraries published before the training date of each LLM. Even so, the M2 low detection performance in Table~\ref{T:va_results} for all the LLMs but GPT-3.5 may designate that these libraries were not considered during LLM training, so the respective scores can be regarded only as indicative. The same applies to the ``I'' score for M2, which it is marked as \textit{N/A}. As discussed in subsection~\ref{SS:LLMs_collection}, to address these limitations, LLMs used for vulnerability detection can capitalize on context augmentation; this way the LLM is provided with access to contextual, up-to-date data.

%\begin{figure}[H]
%	\centering
%	\includegraphics[width=0.95\linewidth]{charts/detection.png}
%	\caption{sample 1 (real numbers) - Excel}
%	\label{F:detection}
%\end{figure}

%\begin{figure}[H]
%	\centering
%	\includegraphics[width=0.95\linewidth]{charts/detection2.png}
%	\caption{sample 2 (Real numbers) - Excel}
%	\label{F:detection2}
%\end{figure}

%\begin{figure}[H]
%	\centering
%	\includegraphics[width=0.95\linewidth]{charts/chart.png}
%	\caption{sample 3 (Demo numbers) - Highcharts}
%	\label{F:detection3}
%\end{figure}

After averaging the ``D'' score for all the nine LLMs, we sort in ascending order the OWASP Top 10 vulnerabilities in Table~\ref{T:va_scoring}. This mean score provides an estimation of the detection difficulty per vulnerability as experienced by the different LLMs. The same table also includes the best performer(s) along with its score in parentheses. As observed from the table, from an LLM viewpoint, M2 is the toughest vulnerability with an average score of 2.11. As explained in subsection~\ref{SS:Evaluation:Process}, this poor outcome is conceivably due to lack of sufficient, up-to-date information at the LLMs' side. Generally, this low score is somewhat expected, as for this vulnerability the LLMs are checking for known security issues in a list of libraries instead of analyzing the application's code. On the other hand, the highest average detection score was observed in M5, where four LLMs achieved a perfect score.

\begin{table}%[H]
\centering
\caption{LLM average detection score per vulnerability. The asterisk exhibitor stands for Code Lama without RAG.}\label{T:va_scoring}
%\begin{adjustbox}{max width=\textwidth}
\begin{tabular}{cp{6cm}c}
\toprule

\textbf{Vulnerability} & \textbf{Best performer} & \textbf{Ave. score} \\ 
\hline
M2 & GPT 3.5 (7) & 2.1 \\ \hline
M1 & GPT 4 (10) & 4 \\ \hline
M7 & Code Llama* (9) & 4.33 \\ \hline
M3 & Zephyr Beta (9), Code Llama* (9) & 4.33 \\ \hline
M8 & Nous Hermes Mixtral (9), Code Llama* (9) & 5.22 \\ \hline
M6 & GPT 4 (10) & 5.33 \\ \hline
M4 & Nous Hermes Mixtral (9) & 5.66 \\ \hline
M9 & MistralOrca (10), Zephyr Beta (10) & 6.77 \\ \hline
M10 & Zephyr Alpha (10) & 7.22 \\ \hline
M5 & Zephyr Alpha (10), Zephyr Beta (10), Llama 2 (10), Code Llama* (10) & 8 \\ 
\bottomrule

\end{tabular}
%\end{adjustbox}
\end{table}

No less important, with reference to the last stage of the experiments as given in subsection~\ref{SS:Evaluation:Process}, regarding the detection of privacy-invasive actions, six, eight, and six of the LLMs correctly perceived potential privacy-invasive actions for location, camera, and local file sharing, respectively. The best performer was Zephyr Alpha, which clearly marked two out of three codes as privacy-invasive and the other as potentially privacy-invasive. The worst performer in this type of experiments was MistralOrca, which was unable to detect any possible privacy-invasive actions.

Additionally, Table~\ref{T:va_results2} presents the results regarding the use of RAG on Code Llama. As explained in subsection~\ref{SS:LLMs_collection}, in this experiment, only half of the samples per vulnerability were indexed for RAG, along with text and code examples from Android's app security guidelines~\cite{AndroidSecGuide} and all the CVEs related to the vulnerable libraries used for M2. After that, we analyzed the other half of the samples, i.e., the non-annotated ones with comments on the particular vulnerability. As observed from Table~\ref{T:va_results2}, the results show improvements in both the detection performance and the generation of code suggestions vis-\`a-vis the base model. Precisely, by feeding a large list of vulnerable libraries, the optimized Code Llama model achieved a perfect score for M2, an improvement of approximately 233\% compared to that in Table~\ref{T:va_results}. Nevertheless, for reaching this performance in real-world scenarios, the RAG process should involve an up-to-date dataset comprising known vulnerable library versions. Interestingly, except M2, the optimized Code Llama model detected the vulnerabilities and suggested improvements for all the M1, M3, M4, M5, M6, and M7 samples. As seen in the three bottom lines of Table~\ref{T:va_results2}, a nearly perfect performance (4/5) was also observed for all the M8, M9, and M10 samples.

\begin{table}%[H]
\centering
\caption{Evaluation results for Code Llama with RAG. The letters ``D'' and ``I'' stand for the number of vulnerable samples detected and the number of vulnerable samples for which the LLM suggested improvements, respectively.}\label{T:va_results2}
%\begin{adjustbox}{max width=\textwidth}
\begin{tabular}{crr}
\toprule
\textbf{Vulnerability} & \textbf{D} & \textbf{I} \\ \hline
M1 & 5/5 & 5/5 \\ \hline
M2 & 10/10 & N/A \\ \hline
M3 & 5/5 & 5/5 \\ \hline
M4 & 5/5 & 5/5 \\ \hline
M5 & 5/5 & 5/5 \\ \hline
M6 & 5/5 & 5/5 \\ \hline
M7 & 5/5 & 5/5 \\ \hline
M8 & 4/5 & 4/5 \\ \hline
M9 & 4/5 & 5/5 \\ \hline
M10 & 4/5 & 5/5 \\
\bottomrule

\end{tabular}
%\end{adjustbox}
\end{table}

As previously mentioned, the performance of the LLMs was also compared against two reputable SASTs, namely Bearer and MobSFscan. Precisely, as shown in Table~\ref{T:va_comparison}, across the 100 samples of Vulcorpus, Bearer found 29 security issues, while MobSFscan detected 12 issues. Excluding M2, this result suggests that, for several vulnerability types, the performance of at least some of the LLMs may significantly or even by far surpass that of well-known SASTs. For instance, comparing the numbers of Table~\ref{T:va_comparison} with the average scores of Table~\ref{T:va_scoring} it can be argued that the former observation applies especially to M3, M4, and M9, and in a smaller extent to M1, M6, and M7. Moreover, a side conclusion is that both the LLMs and SASTs score well in certain vulnerabilities, i.e., M10, and to a lesser extent M5; nevertheless, this is somewhat expected given that vulnerabilities of these two types are generally considered easier to detect.

\begin{table}%[H]
\centering
\caption{Results of prominent SASTs}\label{T:va_comparison}
%\begin{adjustbox}{max width=\textwidth}
\begin{tabular}{cccccccccccc}
\toprule

\textbf{SAST} & \textbf{M1} & \textbf{M2} & \textbf{M3} & \textbf{M4} & \textbf{M5} & \textbf{M6} & \textbf{M7} & \textbf{M8} & \textbf{M9} & \textbf{M10} & \textbf{Total} \\ \hline
Bearer & 2 & N/A & 0 & 1 & 6 & 3 & 3 & 4 & 3 & 7 & \textbf{29} \\ \hline
MobSFscan & 2 & N/A & 0 & 0 & 1 & 1 & 1 & 3 & 0 & 4 & \textbf{12} \\

\bottomrule

\end{tabular}
%\end{adjustbox}
\end{table}

%Last but not least, in terms of privacy-invasive practices, 6, 8, and 6 out of 10 LLMs found possible privacy-invasive actions for location, camera, and local file sharing, respectively. The best performer was Zephyr Alpha, which clearly marked 2 out of 3 codes as privacy-invasive and the other as potentially privacy-invasive. The worst performer was MistralOrca, which was not able to detect any possible privacy-invasive actions.

%No less important, with reference to the last stage of the experiments as given in subsection~\ref{SS:Evaluation:Process}, regarding the detection of privacy-invasive actions, 6, 8, and 6 of the LLMs correctly perceived potential privacy-invasive actions for location, camera, and local file sharing, respectively. The best performer was Zephyr Alpha, which clearly marked 2 out of 3 codes as privacy-invasive and the other as potentially privacy-invasive. The worst performer in this type of experiments was MistralOrca, which was unable to detect any possible privacy-invasive actions.

%\section{Limitations}
%\label{S:Limitations}

%As previously mentioned, it is important to note that the question given to an LLM has a major effect on the output and its ability to detect each vulnerability. Additionally, the performance of RAG implementations is only as good as the dataset given to guide the vulnerability analysis. Another issue is that the size of the data given as context in RAG implementations has an important impact on the analysis time. Therefore a well structure dataset is necessary to reduce hardware requirements.

\section{Conclusions}
\label{S:Conclusions}

Our study provides empirical evidence regarding the effectiveness of using LLMs for Android code vulnerability analysis. GPT-4 and Code Llama emerged as the top performers among the nine LLMs tested, the latter excelling in detection, but failing to provide sufficient code improvements, and the former showing promising results both in detection and code improvement. Notably, the study highlights the superior performance of specific LLMs for particular types of vulnerabilities. For instance, MistralOrca and Zephyr Beta performed exceptionally well for M9, while Zephyr Alpha excelled in M10. These findings suggest that while some LLMs have a general proficiency in vulnerability detection, others may be more specialized, indicating the potential for strategic selection of LLMs based on the targeted vulnerability type. When comparing open LLM models with commercial ones, we can see that the open models were the best performers in seven out of ten categories of vulnerabilities, i.e., M3, M4, M5, M7, M8, M9, M10. On the other hand, considering mean detection and improvements scores, as presented in Table~\ref{T:va_results}, the situation is mixed.
%Open: Nous Hermes Mixtral, llama2, Mistral Orca, Zephyr Alpha, Zephyr Beta, Code Llama
%Commercial: GPT 3.5,4,4 Turbo

Our findings also reveal that while some LLMs are capable of detecting Android code vulnerabilities, their overall performance is still in an early stage. For example, several LLMs struggled with M7, while others were unable to identify M2, reflecting the inherent complexity and subtlety of such vulnerabilities. This outcome points to a need for further research towards enhancing LLMs' capabilities in more nuanced areas of Android security. As an additional step, we evaluated the use of RAG in fine-tuning LLMs for vulnerability analysis, with our results demonstrating that RAG can significantly reinforce detection performance. Regarding the detection of privacy-invasive actions, the obtained results indicate a mixed level of sensitivity among the LLMs, with Zephyr Alpha being the top performer. However, MistralOrca's inability to identify any potential privacy-invasive actions underscores the variability in performance and the need for increased model robustness in privacy analysis concerning mobile platforms. No less important, after comparing the performance of LLMs with that of well-respected SASTs on the same set of vulnerable samples, it can be said that the former seem more adept at identifying code vulnerabilities.

Altogether, the results of the present study provide valuable insights into the current state of LLMs in Android vulnerability detection. While certain models show high efficacy, there is ample room for improvement and targeted optimizations, particularly in addressing complex and subtle vulnerabilities. Nevertheless, for obtaining a more complete view, more experiments with larger datasets are needed.

\bibliographystyle{unsrt}
\bibliography{references}

\end{document}